\documentclass{nature}

\usepackage{amssymb,amsmath,graphicx,longtable}

\usepackage{textcomp}
\def\simlt{\mathrel{\hbox{\rlap{\hbox{\lower4pt\hbox{$\sim$}}}\hbox{$<$}}}}
\def\simgt{\mathrel{\hbox{\rlap{\hbox{\lower4pt\hbox{$\sim$}}}\hbox{$>$}}}}

\def\ale{\mathrel{\hbox{\rlap{\hbox{\lower4pt\hbox{$\sim$}}}\hbox{$<$}}}}
\def\age{\mathrel{\hbox{\rlap{\hbox{\lower4pt\hbox{$\sim$}}}\hbox{$>$}}}}

\begin{document}

\title{A rapid cosmic-ray increase in BC 3372-3371 from ancient buried tree rings in China}

\author{F. Y. Wang$^{1,2}$, H. Yu$^1$, Y. C. Zou$^3$, Z. G. Dai$^{1,2}$ and K. S. Cheng$^4$}

\date{\today}{}
\maketitle

\begin{affiliations}
\item School of Astronomy and Space Science, Nanjing
University, Nanjing 210023, China   %
\item Key Laboratory of
Modern Astronomy and Astrophysics (Nanjing University), Ministry of
Education, Nanjing 210023, China
\item School of Physics, Huazhong University of Science and Technology, Wuhan 430074, China
\item Department of Physics, University of Hong Kong, Hong Kong, China
\end{affiliations}

\begin{abstract}
Cosmic rays interact with the Earth's atmosphere to produce $^{14}$C,
which can be absorbed by trees. Therefore, rapid increases of $^{14}$C
in tree rings can be used to probe
previous cosmic-ray events. By this method, three $^{14}$C rapidly increasing
events have been found. Plausible causes of these events include large
solar proton events, supernovae or short gamma-ray bursts.
However, due to the lack of measurements of $^{14}$C by year,
the occurrence frequency of such $^{14}$C rapidly increasing events is
poorly known. In addition, rapid increases may be hidden in the
IntCal13 data with five-year resolution. Here we report the result
of $^{14}$C measurements using an ancient buried tree during the period
between BC 3388 and 3358. We find a rapid increase of about 9\textperthousand~
in the $^{14}$C content from BC 3372 to BC 3371. We suggest that
this event could originate from a large solar proton event.
\end{abstract}


The cosmogenic nuclide $^{14}$C is produced in the Earth's
atmosphere through neutrons captured by nitrogen nuclei. These
neutrons are generated by cosmic rays interacting with the
atmosphere. Through the global carbon cycle, some of $^{14}$CO$_2$
produced in the atmosphere can be retained in annual tree
rings\cite{Damon95,Damon00,Usoskin06}. Therefore, $^{14}$C
concentrations in annual tree rings indicate the intensity of cosmic
rays. The flux of cosmic rays reaching the earth is modulated by
time variations of geomagnetic and heliomagnetic fields. The
international radiocarbon calibration curve IntCal13 contains tree
ring records of $^{14}$C data with a typical 5-year resolution
extending to 13,900 years before the present\cite{Reimer13}. The
IntCal13 curve shows variations due to solar and geomagnetic
activities on a decadal to millennial time scale. However, there are only a
few annual $^{14}$C data measured from tree rings. So some rapid $^{14}$C
increases caused by cosmic-ray events may be hidden in the IntCal13.

Interestingly, a large amount of cosmic rays can be generated on a
short time scale by high-energy phenomena, such as supernovae
(SNe)\cite{Ackermann13} and large solar proton events
(SPEs)\cite{Miyake12,Melott12,Usoskin13}. Meanwhile, the energy
deposited in $\gamma$-rays of SNe\cite{Damon95} and short gamma-ray
bursts (GRBs)\cite{Meszaros06} can also cause a rapid $^{14}$C
increase. Accordingly, rapid increases of $^{14}$C content in tree
rings are valuable tools to explore high-energy phenomena
occurred in ancient times.

Recently, two events of rapidly increasing  $^{14}$C content occurring
in AD 774-775 and AD 993-994 were found using Japanese
trees\cite{Miyake12,Miyake13}. More recently, a sharp
increase of $^{14}$C content about 10\textperthousand~ within about 3 years
was found at BC 660\cite{Park17}. The AD 774-775 event was
confirmed by other tree samples in different
places\cite{Usoskin13,Jull14,Guttler15}, indicating that this event is worldwide.
Several possible causes for these events have been proposed. Supernova remnants
occurring at AD 774, AD 993 and BC 660 have not been observed and historical
record of SNe has not been found. So the three reported events are unlikely caused by
SNe\cite{Hambaryan13,Chai2015,Park17}. The other possible explanations of
the three events are short GRBs\cite{Hambaryan13,Pavlov13} or large
SPEs\cite{Miyake12,Usoskin13,Jull14,Park17}. The $^{10}$Be measurements in
ice cores from Antarctica, Greenland and Arctic also show a spike
around AD 775\cite{Mekhaldi15,Sigl15}, which indicates that a large
SPE is the most likely explanation. But whether short GRBs could
generate substantial increase in $^{10}$Be remains
unclear\cite{Hambaryan13,Pavlov13}. Meanwhile, the local event rate
of short GRBs is much smaller than the rate of $^{14}$C increase
events. Other attempts have been made to search for rapid $^{14}$C
increases. For example, the $^{14}$C content has been measured in
bristlecone pine tree-ring samples in BC 2479-2455, BC 4055-4031, BC
4465-4441, and BC 4689-4681\cite{Miyake16}. But no large $^{14}$C
increases during these periods are found. It is possible that there
were other rapid $^{14}$C increases in the past, undetected due to the lack of
annual $^{14}$C measurements. In order to find more rapid increases in $^{14}$C data at annual
resolution, we select the periods during which the $^{14}$C value
increases significantly in the IntCal13 data. There are two
intervals where the increase rate is greater than $0.6$\textperthousand ~per year between BC 3380 and 3370. It is possible that
larger annual changes hide in the averaged five-year data.

Here, we report the measurement of $^{14}$C content for an
ancient buried tree in China during the period BC 3388-3358 to
search $^{14}$C increase events, and find a rapid increase from BC
3372 to BC 3371. Considering the occurrence rate of the rapid
$^{14}$C increase events, the $^{14}$C event could originate from a
large SPE.

{\bf Results}

{\bf Wood sample.} We use a Chinese wingnut (Pterocarya stenoptera)
tree, which was found in the city of Yichang, Hubei Province
(30$^{\circ}$31$'$N, 111$^{\circ}$35$'$E), China. The sample of
Chinese wingnut is housed in the Yichang museum. The tree is
shown in Supplementary Figure 1. This type of wood was formed in
the following way. Living trees were buried in river bed or
low-lying place by an earthquake, floods or debris flows. Then,
after thousands of years of carbonization process, this type of wood
would be formed. The carbonization process is the early stage of the
coalification process\cite{Ibarra96}, which can be taken as the slow
hydrothermal carbonization process\cite{Funke10}, and the annual
rings are preserved. The buried wood was first introduced to the
West by Ernest Henry Wilson in 1913\cite{Wilson13}. It is regarded
as priceless raw material for carving. It also has substantial
artistic and scientific research values, such as revealing forest
information, for studying paleoclimate, and speculating on natural
disasters.

Unlike very aging trees, the ancient buried woods are
relatively common in nature. The period of buried woods spans a
long era back. On the other hand, unlike coal, the buried
woods still contain the structure of the trees. Therefore, the
ancient buried woods are good samples of the carbon abundance
research on past epochs up to tens of thousand years ago, as
well as on other plant archaeology. For example, the use of the
sample may extend the data of IntCal13.

The tree was dated with tree-ring records using standard
dendrochronology. We use the master chronology of tree ring widths
from California\cite{noaa}. The program dpLR is used to perform the
dendrochronology\cite{Bunn08}. We find that the correlation value
with the master is 0.525. The possible age error of the wood sample
is about 2 years. Considering the age error of $^{14}$C dating is
about 20 years, the age from dendrochronology is consistent with
that from $^{14}$C dating. We separated annual rings carefully using
a knife. The cellulose samples are prepared by standard chemical
cleaning methods. The tree rings are measured using the Accelerator
Mass Spectrometry (AMS) method at the Beta Analytic radiocarbon
dating laboratory\cite{radiocarbon}. In order to cross check our
results, we also measured another sheet of wood from the same tree
(as a different sample) at the Institute of Accelerator Analysis
laboratory (IAA)\cite{IAA}.

{\bf Measurement data.} In general, AMS measures the fraction of modern carbon $F$, $\delta
^{13}\rm C$ and $\Delta ^{14}\rm C$. The definitions of these values
can be found in Stuiver \& Polach (1977)\cite{Stuiver77}. First we
measure the $^{14}$C content between BC 3388 and 3358 every five
years. Then the yearly measurements of $^{14}$C content from BC 3379
to 3365 are performed, because the $^{14}$C increase rate is greater
than $0.6$\textperthousand~ per year in this period. Figure 1 shows the
variation of $^{14}$C content of the tree rings for the period BC
3388-3358. The filled circles are measured results from the Beta
Analytic radiocarbon dating laboratory and the open circles are
measured from the Institute of Accelerator Analysis laboratory. The
two series of data are consistent with each other (
Table 1 and Table 2). So the measurement results are
reproducible. We find an increase of $^{14}$C content of about
9.4\textperthousand ~from BC 3372 to BC 3371. After the increase, a
gradual decrease over several years due to the carbon cycle is
observed. The significance of this increase with respect to the
measurement errors is 5.2$\sigma$. The profile of this $^{14}$C
event shows a rapid increase within about 1 year followed by a decay
due to the carbon cycle, which is similar to the AD 774-775 event.
In order to estimate $^{14}$C production required for this event, we
use the four-box carbon cycle model to fit the $\Delta^{14}$C data.
The four reservoirs are troposphere, stratosphere, biosphere, and
surface ocean water\cite{Nakamura87}. The transfer coefficients of
carbon from one reservoir to another used in this work are the same
as Miyake et al. (2012). We assume all the $^{14}$C is injected into
troposphere instantaneously. The best-fitting result is shown as the
solid line in Figure 1. The best fit by the weighted least-square
method yields a net $^{14}$C production of
$Q=(7.2\pm1.2)\times10^7\rm\,atoms~cm^{-2}$. According to the
calculation of Usoskin et al. (2013), the $^{14}$C production for
the AD 774-775 event is $(1.3\pm0.2)\times10^8\rm\,atoms~cm^{-2}$.
Therefore, the intensity of this event is about 0.6 times as large
as the AD 774-775 event. In order to compare our results with
IntCal13\cite{Reimer13}, we average the annual data to obtain a
series with 5-year resolution. The result is shown in Figure 2.
Considering the measurement errors, the two sets of data are
consistent with each other. We also compare our data with the
original tree-ring data\cite{Stuiver98,Pearson93} of IntCal13 in
Figure 3. Interestingly, our measured results well agree with the
original data of IntCal13.

{\bf Discussion}

The rapid $^{14}$C increase around BC 3372 must be caused by cosmic
high-energy phenomena. The solar cycle cannot
produce this large increase. There are several plausible origins for this event.

GRBs are the most powerful electromagnetic explosions in the
Universe\cite{Kumar15,Wang15}. According to their duration $T$, they
can be divided into long ($T\geq$2 s) and short ($T<$2 s) GRBs.
Because the intensity of this event is less than that of AD 774-775
event, the energy of a typical short GRB located at a few kpc can
provide necessary energy\cite{Hambaryan13} for this event.
The previous three $^{14}$C events may not be caused by short
GRBs\cite{Mekhaldi15}. So if a short GRB causes this event, it
implies that one short GRB explodes in our galaxy about 5,000
years. But the local rate of short GRBs pointed to the Earth is
$\sim 10^{-5}\,{\rm yr}^{-1}$ \cite{Chai2015,Melott12}. So the short
GRB hypothesis is largely ruled out.

Supernovae are also powerful explosions with high-energy emissions.
For a supernova, the $^{14}$C increase is attributed to both
high-energy photons and cosmic rays, but only the high-energy
photons would be abrupt. Previous work has shown that a rapid $^{14}$C
increase of 6\textperthousand ~occurred three years
after the SN 1006 explosion\cite{Damon95}. However, this result
is challenged by a recent study\cite{Dee17}. The $^{14}$C increase
event reported in this paper occurred about 5,300 years ago, at a time from
which there is no human historical record. Based on the
above calculation, the gamma-ray energy required for this event in
the atmosphere is about $10^{24}$ erg. The typical total energy of a
supernova is $10^{51}$ erg. If a fraction of its total energy,
$\eta_\gamma\simeq 1\%$, radiates in gamma-rays, then the supernova
must be closer than 326 pc. From the Chandra Catalog of Galactic
Supernova Remnants
(http://hea-www.cfa.harvard.edu/ChandraSNR/snrcat\_gal.html), we
find five supernova remnants with distances closer than 400 pc. The
possible ages for these five supernova remnants are: $t=391$ kyr for
G006.4+04.9\cite{Romani10}, $t=4.4\times 10^9$ yr for
G014.7+09.1\cite{Becker99}, $t=3.1\times 10^6$ yr for
G047.3-03.8\cite{Misanovic08}, $t=340$ kyr for
Geminga\cite{Pavlov10}, and $t=2,000-13,000$ yr for G266.2-1.2 (Vela
Jr.)\cite{Allen15}. Indeed, nearby SNe are rare within the last 300
kyr \cite{Melott15}. Interestingly, the Vela Jr.\cite{Aschenbach98}
locates at hundreds of parsecs and its age is $2,000-13,000$
years\cite{Allen15}. From this point of view, a supernova origin of the BC
3372-3371 event appears to be plausible. Unfortunately, $\eta_\gamma$ is much smaller than 1\%\cite{The14},
so a supernova origin for this event becomes highly implausible.

The most probable origin is a large SPE. Usually, SPEs are
associated with solar flares and coronal mass
ejections\cite{Desai16}. Due to the uncertainty of the SPE energy
spectrum, the estimated fluence of SPE caused the AD 774-775 event
varies by as much as two orders of
magnitude\cite{Usoskin13,Thomas13}. Based on more realistic models,
Usoskin et al. (2013) found that the AD 774-775 event could be
explained by a large SPE, which was about 50 times larger than the
SPE in AD 1956\cite{Usoskin13}. So the SPE associated the BC
3372-3371 event is about 30 times larger than the SPE in AD 1956.
According to calculations by Usoskin et al. (2013), the AD 774-775
event required a SPE fluence ($>$30 MeV) of $4.5\times10^{10}$
cm$^{-2}$. Because this event is about 0.6 times as large as the AD
774-775 event, this event should correspond to an SPE fluence about
$2.7\times10^{10}$ cm$^{-2}$, if the same spectrum is assumed. The
SPE fluence above 30 MeV of Carrington event is 1.9$\times10^{10}$
cm$^{-2}$~\cite{Shea06}, which is about 0.7 times as large as that
of BC 3372-3371 event. So rapid increase of radioisotopes should be
detectable around AD 1859. However, neither $^{14}$C nor $^{10}$Be
peaks were found around AD 1859\cite{Miyake13}, which may be
attributed to the different spectra of SPEs. If the BC 3372-3371
event is of solar origin, the associated SPE must be extremely
powerful. In modern society, such extreme events would damage
electronic and power systems\cite{Shea12}, deplete atmospheric
ozone\cite{Lopez-Puertas05} and possibly affect the
weather\cite{Calisto13}. Based on the four events, the
probability of large SPEs is about 7.4$\times 10^{-4}$ yr$^{-1}$.
If we assume the energy of an SPE is comparable to that of an X-ray
flare, the occurrence frequency of large SPEs is consistent
with the frequency of superflares on solar-type
stars\cite{Maehara12}.

In conclusion, we find a rapid increase about 9\textperthousand ~of
$^{14}$C content in buried tree rings from BC 3372 to 3371. Whether
this event is worldwide is unknown. Therefore, measuring the
$^{14}$C content of trees in other places around this period is
important. The most likely origin of this event is a large
SPE. In the future, the
measurements of radiocarbon concentration in tree rings are
important for studying the large SPEs and cosmic gamma-ray events.

{\bf Methods}

{\bf AMS measurement at the Beta Analytic laboratory.}

Cellulose in tree rings is extracted in the following steps: (1)
washing with distilled water; (2) soaking in HCl, NaOH and HCl
solutions; (3) bleaching with hot NaClO$_2$ and washing with boiling
distilled water. Then the material was combusted to CO$_2$ and
converted to graphite using standard procedures. The graphite
powders produced are pressed into AMS targets and measured using an
AMS system at Beta Analytic radiocarbon dating lab in Miami. The
$^{14}$C/$^{13}$C ratio of the sample is compared to known standards
(OxalicI and II, National Institute of Standards and Technology
standards SRM 4990B and 4990C, respectively), and the result
corrected to the measured value of $\delta^{14}$C made off line on a
stable isotope mass spectrometer, giving a value for fraction of
modern carbon.

{\bf AMS measurement at the IAA.}

{\bf Pre-treatment.} (1) Rootlets and granules were removed using
tweezers. (2)  The Acid-Alkali-Acid (AAA) pretreatment process was
used for eliminating carbonates and secondary organic acids. After
the treatment, the sample was neutralized with ultra pure water, and
dried. In the acid treatments of the AAA, the sample is treated with
HCl (1M). In the standard alkaline treatment, the sample is treated
with NaOH, by gradually raising the concentration level from 0.001M
to 1M. If the alkaline concentration reaches 1M during the
treatment, the treatment is described as AAA in the table, while AaA
if the concentration does not reach 1M. (3)  The sample was oxidized
by heating to produce CO$_2$ gas. (4) The produced CO$_2$ gas was
purified in a vacuum line. (5) The purified CO$_2$ gas sample was
reduced to graphite by hydrogen using iron as a catalyst. (6) The
produced graphite was pressed into a target holder with a hole of 1
mm diameter for the AMS $^{14}$C dating, using a hand-press machine.

{\bf Measurement.} The graphite sample was measured against a
standard of Oxalic acid (HOxII) provided by the National Institute
of Standards and Technology (USA), using a $^{14}$C-AMS system based
on the tandem accelerator. A blank for the background check was also
measured.

\noindent \large{\bf Data availability} The data that support the
findings of this study are available from the corresponding author
upon request.

\noindent \large{\bf Acknowledgements} We thank the staff of Yichang
Museum for providing the ancient buried tree and Prof. Kevin Mackeown for critically reading the manuscript. This work is supported
by the National Basic Research Program of China (973 Program, grant
No. 2014CB845800), the National Natural Science Foundation of China
(grants 11422325, 11373022 and 11573014), and the Excellent Youth
Foundation of Jiangsu Province (BK20140016).

\noindent \large{\bf Author Contributions} F.Y.W. conceived the
research. H.Y. prepared samples. All of the authors discussed the
results. F.Y.W. and Y.C.Z. prepared the manuscript. Z.G.D. and
K.S.C. commented on the manuscript. Correspondence and requests for
materials should be addressed to F.Y.W. (fayinwang@nju.edu.cn) or
Y.C.Z. (zouyc@hust.edu.cn).

\noindent \large{\bf Competing interests statement} The authors
declare that they have no competing financial interests.

\newpage

\begin{figure}
\includegraphics[width=\textwidth]{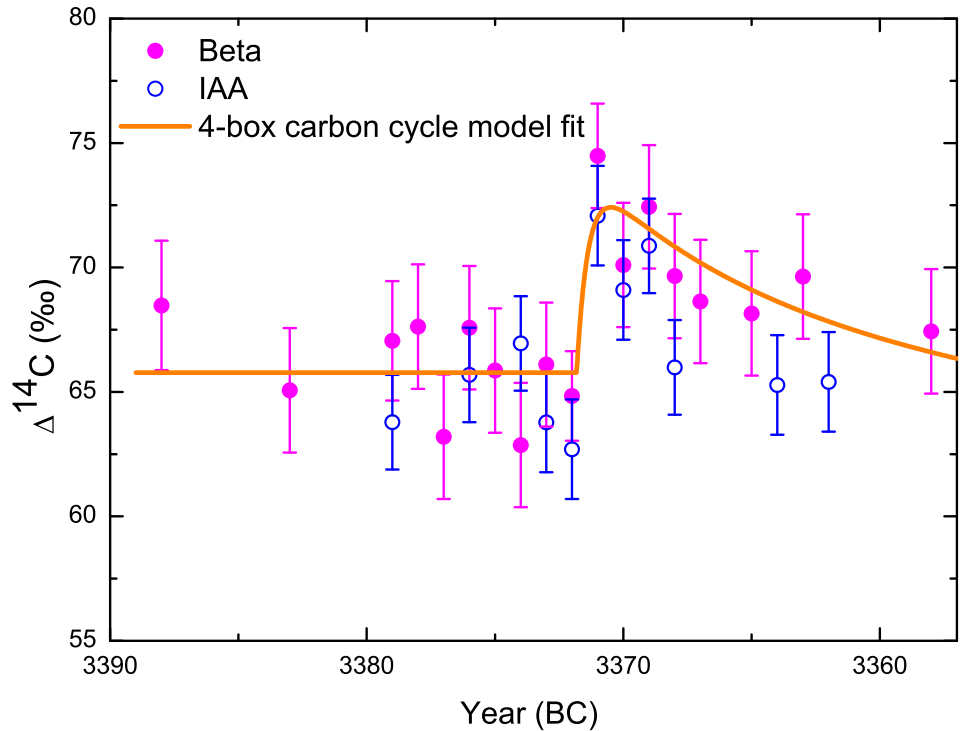}
\caption{\label{Fig1}{\bf Measured $^{14}$C content.} Measured
results of $\Delta$$^{14}$C for the tree rings using the AMS method
at the Beta Analytic radiocarbon dating laboratory (filled circles)
and the Institute of Accelerator Analysis laboratory (open circles).
The typical error of a single measurement is about
2.5\textperthousand ~for filled circles and 2.0\textperthousand ~for
open circles. In order to obtain a smaller error for this $^{14}$C
increase event, several measurements for BC 3371 and 3372 are
performed. The solid line is the best fit for filled circles using
the four-box carbon cycle model with a net $^{14}$C production of
$Q=(7.2\pm1.2)\times10^7\rm\,atoms~cm^{-2}$. Uncertainties (s.d.)
are based on error propagation including measurement errors of the
fraction of modern carbon $F$.}
\end{figure}

\begin{figure}
\includegraphics[width=\textwidth]{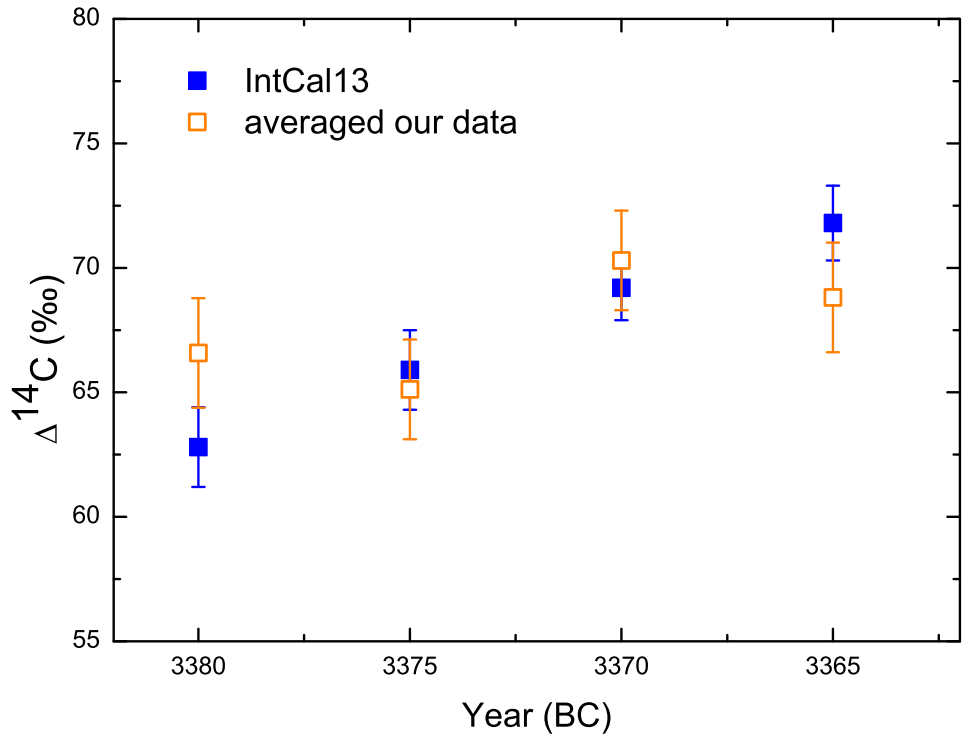}
\caption{\label{Fig2} {\bf Comparison with IntCal13 data.}
Comparison of our five-year average of $\Delta ^{14}$C data measured
in the Beta Analytic laboratory (open squares) with the IntCal13
data (filled squares)\cite{Reimer13}. They are generally consistent
with each other considering the measurement errors. Uncertainties (s.d.) of our data are based on error propagation including
measurement errors of the fraction of modern carbon $F$. The errors
of filled squares are adopted from IntCal13\cite{Reimer13}.}
\end{figure}

\begin{figure}
\includegraphics[width=\textwidth]{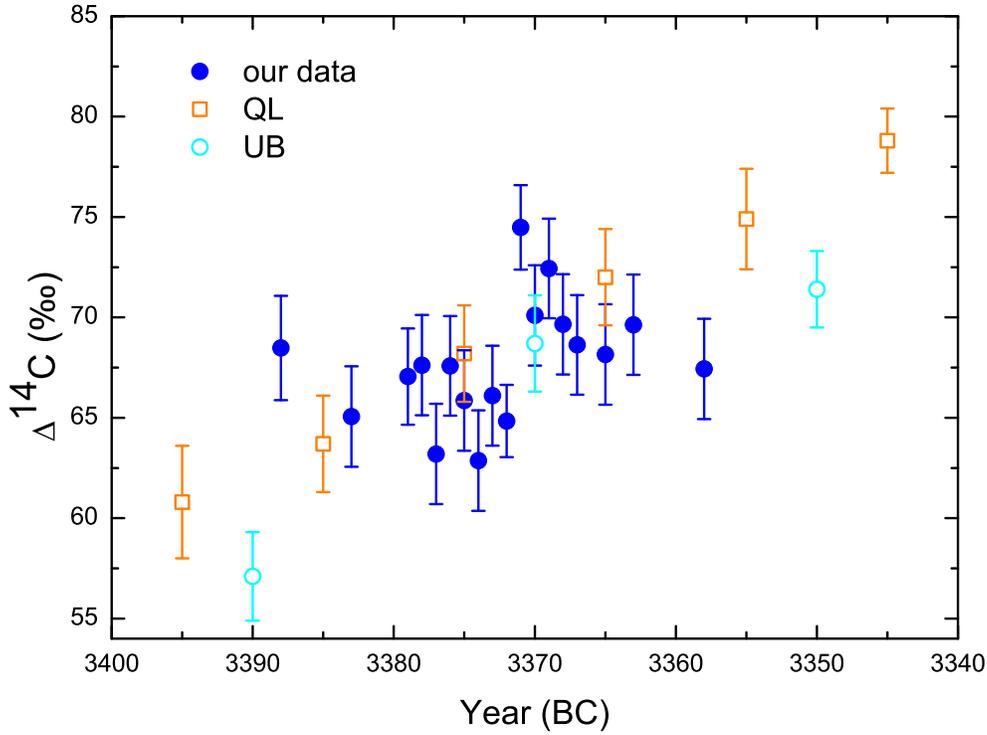}
\caption{\label{Fig3} {\bf Comparison with IntCal13 original data.}
Comparison of our data measured at the Beta laboratory (filled dots)
with the IntCal13 original data [The UW Quaternary Isotope
Laboratory (QL, open squares)\cite{Stuiver98}, The Queen's
University of Belfast (UB, open dots)\cite{Pearson93}]. For the
three $\Delta ^{14}$C value measured in the same year, they are
consistent with each other within measurement errors. The value of
$\Delta ^{14}$C at BC 3385 from QL is between the nearby two points
measured in Beta laboratory. Therefore, our measured results well
agree with the IntCal13 original data considering the measurement
errors. Uncertainties (s.d.) are based on error propagation
including measurement errors of the fraction of modern carbon $F$.}
\end{figure}

\begin{table*}
\centering \caption{Measured results in the Beta Analytic
laboratory. }
\begin{tabular}{ccccc}
\hline\hline Year (BC) & $\Delta^{14}$C~(\textperthousand) & error$^a$ & $\delta^{13}$C~(\textperthousand)\\
\hline
3388  &  68.47 &  2.6 &-24.9\\
3383  &  65.06 &  2.5 &-26.6\\
3379  &  67.05  & 2.4& -26.0\\
3378  &  67.62 & 2.5 & -25.6\\
3377  &  63.19 &  2.5 & -26.1\\
3376  &  67.58 &  2.48  &  -25.2\\
3375  &  65.86 &  2.5 & -26.3\\
3374  & 62.87  &  2.5 & -25.6\\
3373  & 66.10  & 2.49 &  -24.7\\
3372 &  65.03   & 1.8 & -25.4\\
3371  &  74.48 &  2.1 & -24.6\\
3370  & 70.10  & 2.5  & -25.7\\
3369 &  72.44  & 2.48 &  -26.4 \\
3368  & 69.65 & 2.5 & -25.4\\
3367 & 68.63 & 2.48 & -25.6\\
3365 &  68.15 & 2.5 & -25.5\\
3363 & 69.63  & 2.5 & -25.9\\
3358 & 67.43 & 2.5 & -24.8\\
\hline
\end{tabular}\\
$^a$The error of $\Delta^{14}$C is calculated from error
propagation. In the calculation, the error of the fraction of modern
carbon $F$ is considered.
\end{table*}

\begin{table*}
\centering \caption{Measured results in the Institute of Accelerator
Analysis laboratory. }
\begin{tabular}{ccccc}
\hline\hline Year (BC) & $\Delta^{14}$C~(\textperthousand) & error$^a$ & $\delta^{13}$C~(\textperthousand)\\
\hline
3379  &  63.79 &  1.9 &-26.53\\
3376  &  65.68 &  1.9 &-24.77\\
3374  & 66.95  & 1.9 &-25.56\\
3373  &  63.78  & 2.0 &-27.23\\
3372  &  62.70 &2.0 &-25.88\\
3371  &  72.08 &2.0 &-26.16\\
3370  &  69.10 &2.0 &-24.69\\
3369 & 70.87  & 1.9 &-24.97\\
3368 & 65.99  & 1.9& -26.42\\
3364 & 65.28 &2.0 &-25.09\\
3362  &  65.40& 2.0 &-26.04\\
\hline
\end{tabular}\\
$^a$The error of
$\Delta^{14}$C is calculated from error propagation. In the
calculation, the error of the fraction of modern carbon $F$ is
considered.
\end{table*}

\end{document}